\documentclass[12pt,preprint]{aastex}
\received{}
\accepted{}
\slugcomment{ApJ, Vol. 621, March 1, 2005}
\shorttitle{VLBA Imaging of RQQs}
\shortauthors{Ulvestad et al.}
\journalid{}{}
\articleid{}{}
\def\H2{\ion{H}{2}}

\begin{document}

\title{VLBA Imaging of Central Engines in Radio Quiet Quasars}

\author{James S.~Ulvestad}
\affil{National Radio Astronomy Observatory,
P.O. Box O, Socorro, NM 87801} 
\email{julvesta@nrao.edu}
\author{Robert R. J. Antonucci}
\affil{University of California at Santa Barbara,
Dept. of Physics, Santa Barbara, CA 93106}
\email{antonucci@physics.ucsb.edu}
\and
\author{Richard Barvainis\footnote{Any opinions, findings, and conclusions
and recommendations expressed in this material are those of the author and
do not necessarily reflect the views of the National Science Foundation. }}
\affil{National Science Foundation, 4201 Wilson Blvd, Arlington, VA 22230
USA;}
\affil{and Department of Physics, Gettysburg College, Gettysburg, PA 17325}
\email{rbarvai@nsf.gov}

\begin{abstract}

We have used the Very Long Baseline Array (VLBA) to image five
radio-quiet quasars (RQQs) at milliarcsecond resolution, 
at frequencies between 1.4 and 5 GHz.  These quasars have typical
total flux densities of a few millijansky at gigahertz frequencies,
and are compact on arcsecond scales.  The VLBA images reveal that
four of the quasars are dominated by unresolved radio cores, 
while the fifth has an apparent two-sided jet.  Typical core
brightness temperatures range from $10^8$~K to at least $10^9$~K. 
The compact radio morphologies and X-ray luminosities of many
objects in the RQQ sample seem to indicate classical accretion 
onto black holes as massive as $10^9 M_\odot$, with 
emission physics in many ways similar to their radio-loud counterparts.
Therefore, the relatively small amount of radiative energy emerging at
radio wavelengths in the RQQs may simply be due to the presence 
of less powerful radio jets.

\end{abstract}


\section{Introduction}
\label{sec:intro}

Although quasars were first discovered by virtue of their strong radio
emission, most actually fall within the ``radio-quiet'' class,
and thus are known as radio-quiet quasars (hereafter RQQs).  These
objects show optical characteristics similar to 
radio-loud quasars (hereafter RLQs), but ratios of radio/optical \citep{kel89} 
and radio/X-ray emission \citep{ter03} are one to several orders of magnitude
below the RLQs.  RLQs and RQQs often have been thought to be distinct 
populations \citep{kel89,hoo95}, although the gap may be
bridged by a significant population of radio-intermediate
objects \citep{whi00,ho01}.  Arcsecond-scale
radio emission from RQQs sometimes appears similar
to the RLQs, including the presence of relatively large
double/jet radio sources \citep{kel94,kuk98,blu01}.  

Regardless of the existence (or not) of two separate quasar
classes, it is true that the radio/X-ray or
radio/optical ratio can differ by several orders of magnitude
between RQQs and RLQs.  Possible reasons for this
wide disparity in radio power between the two classes have been
discussed for a number of years without a clear consensus.
Some RQQs show a high-frequency excess, with apparent
flat or inverted radio spectra at gigahertz frequencies 
(Antonucci \& Barvainis 1988; Barvainis, Lonsdale, \& Antonucci 1996;
Barvainis \& Lonsdale 1997).  It has been speculated
that this emission might include a component due to 
thermal radio emission, and that this component might be
related to the other differences in quasar properties.  
The thermal emission could be at $10^4$~K, or even at a
brightness temperature above $10^6$~K, as has been inferred in
the Seyfert galaxy NGC~1068 \citep{gal97,gal04}.  However, recent work
on radio variability of RQQs (Barvainis et al.\ 2004)
suggests that in many cases flat or 
inverted spectra in RQQs are a result of partially opaque synchrotron
cores like those found in RLQs.   
Milliarcsecond-resolution radio imaging can provide an independent
test of the emission physics of RQQs and may illuminate any
key differences between RQQs and RLQs.  The brightness temperature 
corresponding to a few-millijansky source unresolved by the 
Very Long Baseline Array (VLBA) is
$\sim 10^8$~K, roughly independent of frequency. Therefore, the
VLBI technique provides a good filter for the presence of nonthermal 
radio emission.

Only a few RQQs have been imaged at milliarcsecond resolution.  
\citet{blu96} first detected a milliarcsecond core in the RQQ E1821+643,
and high-brightness-temperature cores later were found in a few 
more objects by \citet{blu98} and \citet{cac01}.  
The typical brightness temperatures above $10^8$~K in these cases 
apparently ruled out thermal emission from a disk or torus.  In
addition, observations by Blundell, Beasley, \& Bicknell (2003) 
suggest a brightness temperature high
enough to require superluminal motion in one RQQ. 
The present paper reports further milliarcsecond 
imaging of a sample of five RQQs, including multi-frequency
imaging of several objects, using the VLBA.  The goal was to determine
whether the RQQ radio emission is dominated by nonthermal
processes, or whether any RQQs might have radio properties
affected by emission or absorption from a thermal gas.

\section{Sample Selection}

In order to assess the milliarcsecond structure of 
RQQs chosen by some well-defined criteria, we began with the sample
listed by \citet{bar96} and \citet{bar97}. From their lists, we
selected the 11 RQQs having total 5-GHz flux densities of 4~mJy or 
greater.  Three of these objects show no cores stronger than 1~mJy
in VLA images \citep{kuk98}, so we discarded them from the sample.
Eight RQQs remained, four at relatively low redshift ($z\leq 0.33$)
and four at considerably higher redshift ($0.94 < z < 2.4$).
Properties of these RQQs are listed in Table~\ref{tab:sample}.  
Our VLBI RQQ sample is not complete in
any sense, but is subject to the heterogeneity present in the
original selection by \citet{bar96} and \citet{bar97}.

\section{Observations and Imaging}

We observed six of the eight members of our sample in 1999 and 2000, 
using the VLBA, which is described by \citet{nap93}; the two other
RQQs had been observed previously with the VLBA \citep{blu98,blu03}. 
J1219+0638 and J1353+6345 also were imaged by \citet{blu98}, 
but we observed them at multiple frequencies to derive spectral
information useful for constraining their emission processes.
Table~\ref{tab:obs} summarizes the VLBA observations.  
In most cases, all 10 VLBA antennas participated successfully.  
However, data and/or antennas occasionally were removed due
to snowstorms, radio-frequency interference, or instrumentation
failures.

All target quasars were below the flux-density threshold needed
to solve for instrumental and atmospheric effects,
so each was phase-referenced
\citep{bea95} to a nearby strong compact radio source.   
Phase-reference cycle times were 4--5 minutes, and total on-source 
times ranged from 1 to 4 hours per frequency band.  Charged-particle
effects were corrected with global ionospheric models\footnote{Ionospheric
models are available from {\it cddis.gsfc.nasa.gov.}} using AIPS, the 
Astronomical Image Processing System \citep{gre03}.  Amplitudes were
calibrated using standard gain files as well as system temperatures
measured at 1--2 minute intervals, then checked by observations of 
simple strong sources.  Global clock offsets were found from observations 
of strong sources, while atmospheric and electronic drifts were 
calibrated using the local phase-reference sources.  
Imaging of J1316+0051 failed, since the distance of 
$8^\circ$ from the phase-referencing source prevented
successful atmospheric calibration.  The other 
quasars had phase-reference sources between $1.5^\circ$
and $4.5^\circ$ from the target.  Although their peak flux densities 
initially were reduced by imperfect atmospheric 
calibration, all had enough signal for 
self-calibration, which largely eliminated the image degradation.

Final images were made using two different data weightings,
pure ``natural'' weighting which maximizes the sensitivity,
and a compromise between natural and ``uniform'' weighting,
which provides a good combination of sensitivity and
resolution.  J1219+0638, J1353+6345, and J1436+5847  
were observed at multiple frequencies
and are unresolved at all bands, while
J0046+0104 also was unresolved at its only observed 
frequency of 4.99 GHz.  J0804+6459 contains
a core with weak extended emission at 4.99 GHz, but a prominent
jet at 1.67 GHz.  The 4.99 GHz VLBA images of the quasars 
are shown in Figure~1, and a 1.67 GHz image also is displayed 
for J0804+6459.

Table~\ref{tab:tb} lists radio positions, flux densities,
powers, and brightness temperatures for each quasar.
For all except J0804+6459, these results come from Gaussian fits
to the core components, constrained to be completely unresolved by 
the synthesized beams. 
For J0804+6459, the total flux densities were found by integrating
over the significant emission in the image.  
Source positions have typical errors of 1~mas at 4.99~GHz, 
dominated by the errors in the reference source positions and 
generally independent of angular separation between the target
and reference sources.  Source positions at the lower frequencies 
have larger errors due to the Earth's ionosphere, and therefore are 
not listed.  Radio powers in the source frames that were emitted 
at our observed frequencies were computed from the flux densities 
using Ned Wright's cosmology calculator.\footnote{The on-line
cosmology calculator may be found at 
{\it http://www.astro.ucla.edu/$\sim$wright/CosmoCalc.html}}
A flat cosmology was assumed, with
$H_0=71$~km~s$^{-1}$~Mpc$^{-1}$, $\Omega_m=0.27$, and 
$\Omega_\Lambda=0.73$ \citep{spe03}.
The K-correction was computed using the
spectral indices between the two most widely spread VLBA frequencies
when we observed at multiple frequencies, or using the VLA spectral
index from Table~\ref{tab:sample} for J0046+0104. 
Peak brightness temperatures 
were computed in the source rest frames, using the formula
given by \citet{fal96b}, modified for an elliptical Gaussian 
and multiplied by $(1+z)$ to account for the source redshift:
\begin{equation}
T_{\rm B}= 1.8\times 10^9 (1+z) \biggl(\frac{S_{\nu}}{\mathrm{mJy}}\biggr)
\biggl(\frac{\nu}{\mathrm{GHz}}\biggr)^{-2} 
\biggl({\theta_1 \theta_2\over {\rm mas}^2}\biggr)^{-1} \mathrm{K}\ .
\label{eqno1}
\end{equation}
Here, $S_\nu$ is the flux density at frequency $\nu$, with $\theta_1$
and $\theta_2$ being the fitted full widths at half maximum of the
major and minor axes of the sources.  For the unresolved sources, 
the upper limits to the component sizes are taken to be one-half 
the beam sizes; the synthesized beam size is used for the 
brightness temperature in J0804+6459, since its peak flux density
decreases systematically with increasing resolution, and there
is no conclusive evidence for a completely unresolved component.

\section{Discussion}

\subsection{Individual Source Properties}

\subsubsection{J0046+0104}

J0046+0104 (UM 275) is a broad absorption line quasar at a redshift 
of $z=2.137$.  \citet{shi03} report a central
black hole mass of $2\times 10^9 M_\odot$, based on H$\beta$ line
widths.  The single radio component is unresolved, with a flux density
of 4.7~mJy, consistent with published VLA results.
The observed 0.2--3.5~keV flux upper limit is
$<4.1\times 10^{-13}$~ergs~cm$^{-2}$~s$^{-1}$\ \citep{wil94}.
We convert to a 2--10~keV limit and compute
the radio/X-ray flux ratio $R_X\equiv \nu S_\nu({\rm 5\ GHz})/
F_X({\rm 2-10\ keV})$ defined by \citet{ter03}, finding
$R_X > 1\times 10^{-3}$.  This is in the realm 
occupied by radio-loud objects according to \citet{ter03}.

\subsubsection{J0804+6459}

J0804+6459 also is a broad absorption line quasar, at $z=0.148$, 
having a much less powerful radio source than the more distant 
J0046+0104.  It is a relatively strong IRAS source; a {\it Hubble
Space Telescope} image of the quasar and host galaxy is shown by
\citet{boy96}.  As might be expected from its steep radio spectrum, 
J0804+6459 is significantly resolved by the VLBA.  The position
angle of its apparent two-sided radio jet changes by $\sim 30^\circ$
between the smallest and largest scales sampled by our VLBA
observations, and the maximum radio source size of $\sim 80$~mas
corresponds to $\sim 200$~pc.  
The 0.5--8.0~keV flux density observed by {\it Chandra}
is $4.2\times 10^{-14}$~ergs~cm$^{-2}$~s$^{-1}$ \citep{gre01}.
Converting this to a 2--10~keV flux and using the peak VLBA
flux density of 2.4~mJy at 5~GHz, we find
$R_X\approx 5\times 10^{-3}$, firmly within the class of 
objects with radio-loud cores.

\subsubsection{J1219+0638}

\citet{blu98} found ``structures indicative of jets'' in J1219+0638,
using their short VLBA observations at 8.4~GHz.  However, our
deeper imaging at three lower frequencies, where such jets would be
more prominent, reveals nothing except an unresolved
point source.  Our measured 5-GHz flux density of
$6.4\pm 0.3$~mJy differs from published VLA values of 
$9.1\pm 0.5$~mJy \citep{bar96} and
$4.0\pm 0.1$ \citep{kel89}.  This apparent radio
variability and the overall flat spectrum support the
result that the RQQ is, in fact, completely unresolved
by the VLBA.  The
rest-frame brightness temperature of $> 5\times 10^8$~K, 
indicates that free-free absorption or emission are unlikely
to account for the spectral shape.  Instead, this RQQ almost
surely contains a synchrotron-self-absorbed radio core.  We
note that a typical brightness temperature for a self-absorbed source
is $\sim 10^{10}$~K or higher.  This implies that the actual source
diameter may be at least a factor of five smaller than our upper
limit, or smaller than 0.2--0.3~mas ($\sim 1$~pc at $z=0.334$).  

J1219+0638 is a relatively strong X-ray source, with an observed 
0.1--2.4~keV flux of $5.2\times 10^{-12}$~ergs cm$^{-2}$ s$^{-1}$ 
(Brinkmann, Yuan, \& Siebert 1997),
about two orders of magnitude higher than J0804+6459, despite its
somewhat greater distance.  Therefore, the radio/X-ray ratio of 
$1\times 10^{-4}$ is considerably smaller, and
J1219+0638 is near the boundary between radio-loud and radio-quiet
objects. 

\subsubsection{J1353+6345}

Surprisingly, we find no hint of resolution in the steep-radio-spectrum
quasar J1353+6345 ($z=0.087$),
again contradicting the inference of \citet{blu98}.
Our 4.99~GHz flux density is consistent with the
VLA measurement by \citet{bar96}, though well below a measurement
by \citet{kel89}, and the 1.42~GHz flux density is nearly a
factor of two less than found by \citet{bar96}.  The brightness temperatures
for the unresolved core approach the values where we might
expect synchrotron-self-absorption to occur, yet we find no
evidence for such absorption in the radio spectrum.  Unlike J1219+0638,
we therefore suspect that the angular size of the radio source in
this RQQ is not much smaller than our upper limit at 4.99~GHz,
and is likely to be $\sim 0.5$~mas ($\sim 0.8$~pc).
The observed 0.1--2.4~keV flux is $1.6\times 10^{-12}$~ergs~cm$^{-2}$~s$^{-1}$
\citep{bri97}, implying $R_X\approx 4\times 10^{-4}$, within the radio-loud
category.

\subsubsection{J1436+5847}

J1436+5847 (Mrk817, or UGC 09412) is the most nearby object in our sample,
at $z=0.033$, and sometimes has been classified as a Seyfert 1 galaxy
rather than a quasar.  Its radio power of $\sim 10^{22}$~W~Hz$^{-1}$
is within the upper end of the range found for nearby classical
Seyfert galaxies \citep{ulv89} and Palomar Seyfert galaxies
\citep{ho01b,ulv01}.
Previous VLA observations \citep{ulv84,bar96} give consistent
flux densities of 5~mJy at 5~GHz, slightly higher than the 
value of $3.1\pm 0.2$~mJy found here for the VLBI core.
J1436+5847 is another steep-spectrum source that is unresolved
by the VLBA, and appears to be a somewhat less powerful
``twin'' of J1353+6345 at radio frequencies.

The 0.2--2.4~keV flux observed by ROSAT is 
$4.0\times 10^{-12}$~ergs~cm$^{-2}$~s$^{-1}$ \citep{pri02}.
Converting the soft energy flux to a higher energy band, we 
find $R_X\approx 6\times 10^{-5}$, putting the quasar
near the border between radio-loud and radio-quiet objects.

\subsection{Source Brightness Temperatures}

Seven of the eight RQQs in the sample listed in Table~\ref{tab:sample}
now have been imaged with the VLBA, so we may say something about
their global properties.  
These objects actually have a very wide range of radio powers at
5~GHz, as expected from their similar flux densities and wide
range of redshifts, yet their brightness temperature values or limits are 
remarkably similar.  The actual values/limits are subject to a strong
selection effect, namely the fact that sources with a few millijansky
of compact flux detected on scales of one to a few milliarcseconds inevitably
must have brightness temperatures of $10^8$~K or greater.
Still, it is a key result that essentially all the RQQs, including those
with flat and steep spectra, have some components with minimum brightness 
temperatures of $\gtrsim 10^8$~K (cf. Table~\ref{tab:tb}).  
This implies that all have nonthermal emitting cores, rather than 
being dominated by thermal tori such as the one inferred for NGC~1068 
by Gallimore et al. (1997, 2004).  Further, the four objects
that we observed at multiple frequencies show no evidence for
low-frequency radio turnovers, as shown in Figure~\ref{fig:spect}.
This implies that free-free absorption at $\sim 1.5$~GHz is not
significant in most RQQs, and thus cannot easily account for 
flat radio spectra at gigahertz frequencies.

\subsection{Generation of the Radio Emission}

Why do RQQs contain such compact, and yet relatively low-luminosity
(compared to RLQs) radio emission?  A likely possibility, of course,
is that this weak emission is somehow related to the process of accretion 
on supermassive black holes at the centers of the quasar host galaxies,
and to the efficiency of radiation from that accretion process.
A key element to consider here is the overall spectral energy 
distributions of the RQQs.  Figure~\ref{fig:sed} plots the radio/X-ray
ratio versus the radio/optical ratio for the RQQs in our sample. 
The radio/optical ratio, $R_O$, is defined
as the ratio of observed 5 GHz and $B$ band flux densities, where
the conversion from $B$ magnitude to flux density is given by
\citet{sch83}.  We computed the radio/X-ray ratio, $R_X$, converting 
from the observed energy bands to 2--10~keV fluxes using 
$N(E)\propto E^{-2}$. This typically reduces the X-ray fluxes by a
factor of order 2; given this fact and the lack of knowledge of
the spectra,
$R_X$ values may be in error by factors of 2--3.  For the 
five RQQs that we observed with the VLBA, the 5-GHz radio power 
from the unresolved milliarcsecond core has been used;
otherwise we use the total VLA flux density (Table~\ref{tab:sample}).
Diagonal lines are plotted corresponding to O/X ratios of 1 and 100,
where this ratio is defined as the ratio of $\nu S_\nu$ at 4400\AA\ to
the 2--10~keV flux.  The radio-loud quasars
3C~273 and 3C~454.3 also are plotted, with the two points corresponding to 
their total 5~GHz flux densities and the flux densities of their
unresolved VLBI cores \citep{unw85,hom01}.

Figure~\ref{fig:sed} shows some correlation between $R_O$ and $R_X$, 
as expected since the radio flux density appears on both axes.
Most of the eight RQQs in our sample have
$1\lesssim R_O \lesssim 30$, which might be considered to indicate
``radio-intermediate'' quasars (e.g., Falcke, Sherwood, \& Patnaik 1996),
and all the RQQs have $\log(R_X) > -4.5$, fitting the definition
of radio-loud objects \citep{ter03}.  Of course, our choice of 
objects that could be observed with VLBI leads to a clear selection 
effect favoring relatively stronger radio sources.
Since Figure~\ref{fig:sed} shows that $R_X$ may differ by one
to two orders of magnitude at a given value of $R_O$, the implication
is that the distribution of accretion energy into radiated power is 
very different for the individual objects.  For the two RLQs plotted,
the optical/X-ray ratios are somewhat lower than for the RQQs, but the
strong variability of 3C~273 and 3C~454.3 makes it difficult to conclude
that this might be a general trend.  If it is, it could indicate
an effect such as additional inverse-Compton X-ray emission from strong
compact VLBI jets.

We can assess further the accretion process in RQQs by taking
the RQQ J1409+2618 as an
example.  For a black hole mass of $\sim 10^9M_\odot$,
its Eddington luminosity would be 
$\sim 1.4\times 10^{47}$~ergs~s$^{-1}$.  The
$B$ magnitude of 16.07 corresponds to an optical ``luminosity''
$\nu L_B\approx 5\times 10^{46}$~ergs~s$^{-1}$, while the X-ray
luminosity is considerably weaker.  Thus the bolometric luminosity 
is probably close to the Eddington luminosity.  
Analysis for the other high-redshift RQQs in our sample gives similar
results, indicating classical accretion on $\sim 10^9M_\odot$ black holes,
consistent with results found by \citet{ves04}.  For our four RQQs 
with $z\leq 0.33$, the lower luminosities may
indicate either low-radiative-efficiency accretion \citep{nar98} or 
lower mass black holes.  A number
of optically selected quasars having $z\sim$0.1--0.2 seem to contain
black holes of only $10^7$--$10^8M_\odot$ \citep{shi03}, favoring
the latter possibility for our nearby RQQs.

Based on the above arguments and the observed brightness temperatures,
we rule out either thermal processes or low-efficiency accretion
for most RQQs.  Why, then, are they different from radio-loud objects?
One possibility could be that the radio emission in RQQs
is not beamed along our line of sight.
Such simple orientation schemes are ruled out, however,
by the fact that RQQs do not show bright, large-scale radio structures 
(arcsecond- to arcminute-scale jets or double lobes) nearly as
powerful as those in RLQs.  An alternative scenario would have RQQ radio 
cores fundamentally similar to RLQs, with the RQQ jets 
simply being less powerful.   Following Blundell \& Rawlings (2001)
and \citet{bar04}, RQQs may produce diffuse, Fanaroff-Riley type I 
\citep{fan74} large-scale structures as their weak jets are disrupted before 
escaping their host galaxies.   In at least one RQQ, such a structure does 
exist \citep{blu01}, but only very deep integrations are capable 
of detecting these structures.

The most definitive test of a connection between RQQs and RLQs 
would be to achieve much more sensitive VLBI imaging in order 
to determine whether milliarcsecond jets are present in most RQQs, and 
eventually to measure the jet speeds.  \citet{blu03} present 
evidence for a Doppler factor approaching 10 for a VLBI component 
in the RQQ J1409+2618, one of our sample members.
Direct measurement of superluminal motion, however,
has yet to be achieved; we plan to 
search for superluminal jets in RQQs using the new High 
Sensitivity Array of VLBI stations.

\section{Summary}

We have imaged five RQQs at milliarcsecond resolution using the
VLBA.  Four of these objects are unresolved, while a fifth shows
an apparent two-sided radio jet.  The brightness temperatures and
spectra of the radio cores of these RQQs appear to rule out
the possibility that thermal emission or absorption can be
important contributors to their properties.  Although the
radio emission is relatively weak compared to radio-loud
quasars and blazars, the overall spectral energy distributions
and variability properties seem to indicate classical accretion 
onto black holes
as massive as $10^9 M_\odot$ for the more distant objects.
The radio/optical ratios are fairly low, but the radio/X-ray
ratios of many of our sample members actually are consistent
with values for radio-loud objects.  Thus, the most likely 
explanation for their radio properties is simply that the RQQs 
are similar to their ``traditional'' radio-loud 
cousins, but with less powerful radio jets.

\acknowledgments

The National Radio Astronomy Observatory is a facility of the 
National Science Foundation operated under cooperative agreement by Associated 
Universities, Inc.  We thank the staff of the VLBA for carrying out these
observations in their usual efficient manner.  This research has
made use of the NASA/IPAC Extragalactic Database (NED) which is 
operated by the Jet Propulsion Laboratory, California Institute of 
Technology, under contract with the National Aeronautics and Space
Administration; it also has made use of NASA's Astrophysics Data
System Bibliographic Services, and Ned Wright's on-line cosmology 
calculator.  We thank the referee for suggestions that helped
considerably in clarifying the paper.


\clearpage

\begin{figure}
\plottwo{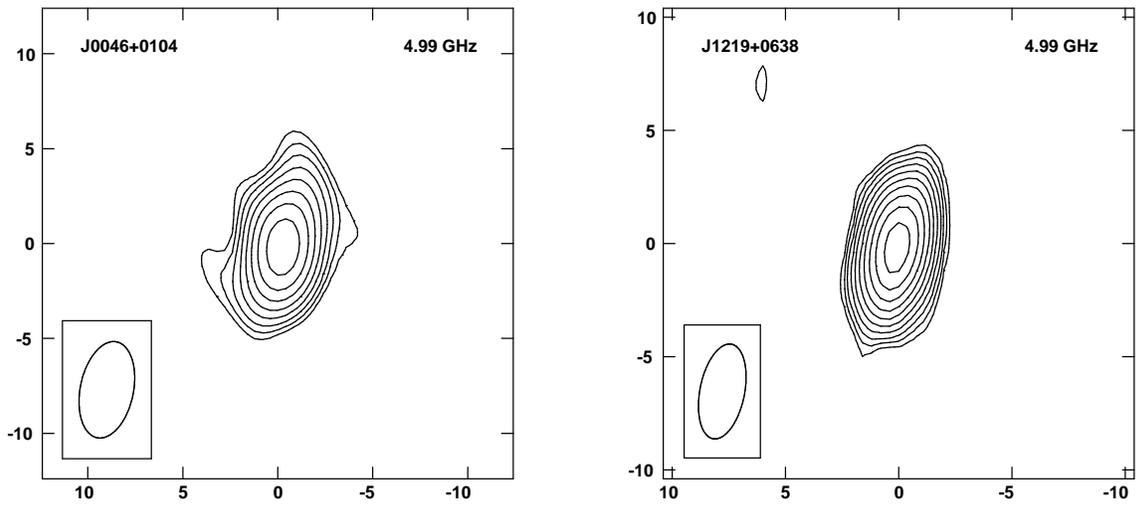}{f1d.eps}
\caption{
VLBA images of 5 RQQs imaged in our program, all in left circular
polarization at 1.67 or 4.99 GHz.  The
quasar names and observing frequencies are given inside each individual
panel, with axes labeled in milliarcseconds referenced to 
the centers of the displayed images.  
All images are naturally weighted, with the contour levels
separated by factors of $\sqrt{2}$, beginning at 3 times the rms
noise given in Table~\ref{tab:obs}.  Negative contours are shown
dashed, and half-power beam sizes are shown in the lower left corners.}
\label{fig:vlba-rqq}
\end{figure}

\clearpage

\setcounter{figure}{0}

\begin{figure}
\plottwo{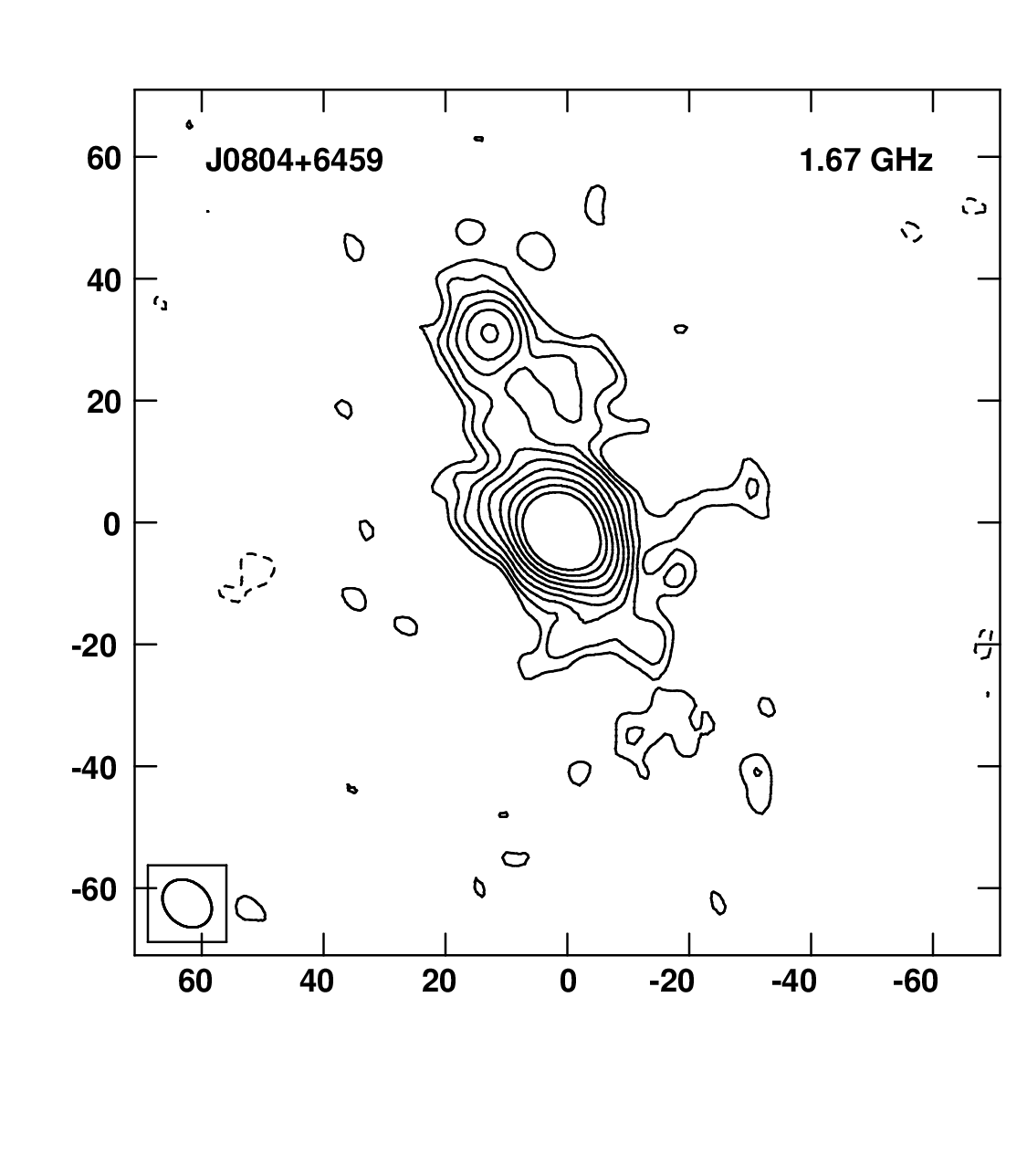}{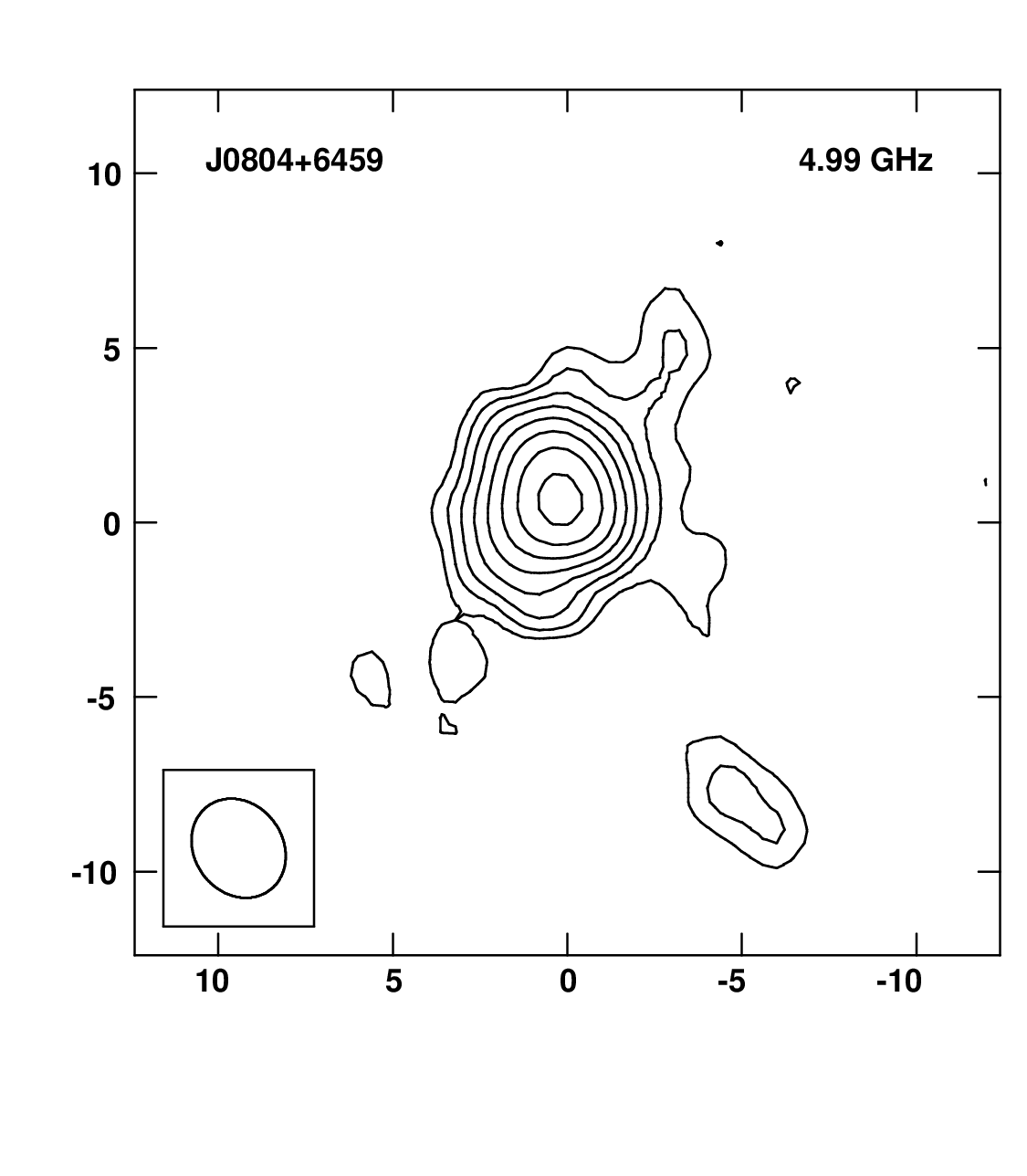}
\caption{
VLBA images (continued).}
\end{figure}

\clearpage

\setcounter{figure}{0}

\begin{figure}
\plottwo{f1e.eps}{f1f.eps}
\caption{
VLBA images (continued).}
\end{figure}

\clearpage

\begin{figure}
\plotone{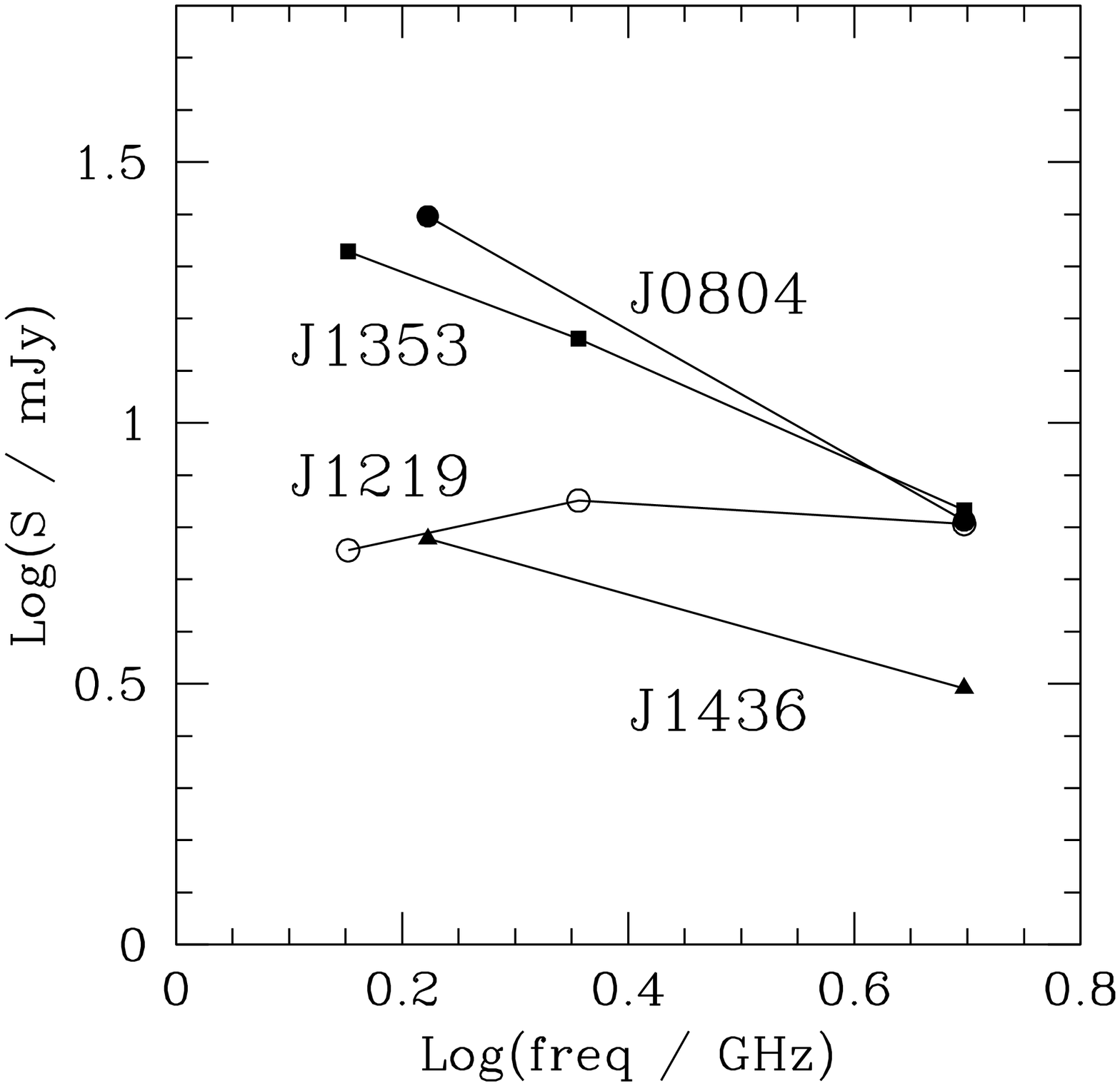}
\caption{
Radio spectra of the VLBI sources in four different
RQQs imaged for this paper.  Typical flux-density errors 
are 0.03 dex, about the size of the plotted points.}
\label{fig:spect}
\end{figure}

\clearpage

\begin{figure}
\plotone{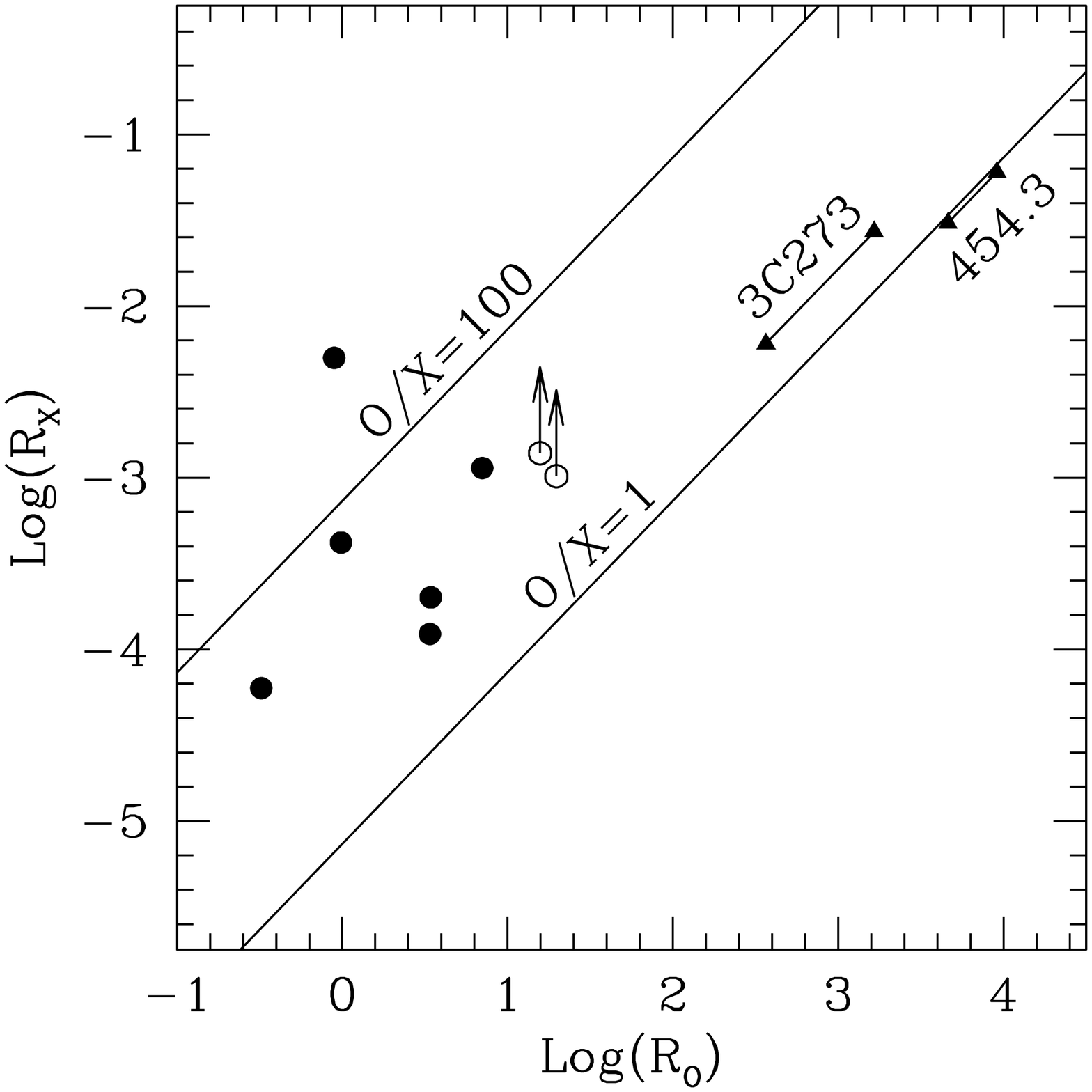}
\caption{
The radio/X-ray ratio, $R_X$, is plotted against the radio/optical
ratio, $R_O$, for the eight RQQs in our sample.  Both
$R_O$ and $R_X$ are defined in the text.  Closed circles represent 
objects detected at X-ray, optical, and radio wavebands, while the open circles
with arrows represent the two RQQs that are not detected at X-ray 
wavelengths.  Errors in $R_X$ may be quite large, perhaps as high as 0.5 dex 
in individual cases.  Diagonal lines indicate optical/X-ray ratios of
1 and 100, where $O/X=\nu S_\nu$ (4400 \AA)/$F_X$(2--10 keV).
The radio-loud quasars 3C~273 and 3C~454.3 also are plotted, with two values
corresponding to total and unresolved (by VLBI) radio flux densities.}
\label{fig:sed}
\end{figure}

\clearpage

\begin{deluxetable}{lccccccc}
\tabletypesize{\scriptsize}
\tablecolumns{8}
\tablewidth{0pc}
\tablecaption{Objects in RQQ Sample}
\tablehead{
\colhead{(1)}& \colhead{(2)}& \colhead{(3)}& \colhead{(4)}&
\colhead{(5)}& \colhead{(6)}& \colhead{(7)}& \colhead{(8)} \\
\colhead{Name}&\colhead{$z$\tablenotemark{a}}&\colhead{$B$\tablenotemark{a}}&
\colhead{$S$(5 GHz)\tablenotemark{b}}&\colhead{$\alpha$(1.4,5)\tablenotemark{b}}&
\colhead{$F_X$(2--10 keV)\tablenotemark{c}}&\colhead{X-ray Band\tablenotemark{c}}&
\colhead{X-ray Ref.\tablenotemark{d}} \\
\colhead{}&\colhead{}&\colhead{(Mag.)}&\colhead{(mJy)}&\colhead{}&
\colhead{(ergs cm$^{-2}$ s$^{-1}$)}&\colhead{(keV)}&\colhead{}}
\startdata
J0046+0104\tablenotemark{e}&2.137&(18.2)&4.1&0.0&$<2.3\times 10^{-13}$&
0.2--3.5&1 \\
J0804+6459\tablenotemark{e}&0.148&(14.9)&11.7&$-$1.1&$2.4\times 10^{-14}$&
0.5--8.0&2 \\
J1219+0638\tablenotemark{e}&0.334&(15.95)&9.1&+0.5&$2.6\times 10^{-12}$&
0.1--2.4&3 \\
J1225+2235&2.058&16.6&7.3&0.0&$3.1\times 10^{-13}$&0.2--3.5&1 \\
J1316+0051&2.393&18.13&4.0&+0.1&$<1.4\times 10^{-13}$&0.1--2.4&4 \\
J1353+6345\tablenotemark{e}&0.087&14.54&7.0&$-$1.5&$8.1\times 10^{-13}$&
0.1--2.4&3 \\
J1409+2618&0.945&16.07&5.8&$-$0.3&$1.4\times 10^{-12}$&0.5--10&5 \\
J1436+5847\tablenotemark{e}&0.033&14.19&5.1&$-$1.0&$2.6\times 10^{-12}$&
0.2--2.4&6 \\
\enddata
\tablenotetext{a}{All redshifts and $B$ magnitudes are taken 
from the on-line version of \citet{ver03}.  Magnitudes given in
parentheses are derived from $V$ magnitudes given by \citet{ver03},
assuming $B-V=0.3$, or from the NASA Extragalactic Database (NED).}
\tablenotetext{b}{The 5-GHz flux densities and spectral indices between
1.4 and 5 GHz are taken from the VLA observations of \citet{bar96}
and \citet{bar97}.}
\tablenotetext{c}{The 2--10 keV X-ray fluxes in Column (6) are computed
from values derived for the energy range given in Column (7), assuming
that $N(E)\propto E^{-2}$.}
\tablenotetext{d}{References for X-ray fluxes: 1. \citet{wil94}.
2. \citet{gre01}. 3. Brinkmann, Yuan, \& Siebert (1997). 4. \citet{vog99}. 5. \citet{ree00}. 6. \citet{pri02}.}
\tablenotetext{e}{Object observed successfully by the VLBA for the
present paper.  See additional discussion in text as well as
information in Table~\ref{tab:obs}.}
\label{tab:sample}
\end{deluxetable}

\clearpage

\begin{deluxetable}{lccccccc}
\tabletypesize{\scriptsize}
\tablecolumns{8}
\tablewidth{0pc}
\tablecaption{VLBA Observations}
\tablehead{
\colhead{(1)}& \colhead{(2)}& \colhead{(3)}& \colhead{(4)}&
\colhead{(5)}&\colhead{(6)}& \colhead{(7)}& \colhead{(8)} \\
\colhead{Name}&\colhead{Date}&\colhead{Ref. Source}& \colhead{Frequency}& 
\colhead{Pol.\tablenotemark{a}}&\colhead{Integration}&\colhead{rms\tablenotemark{b}} 
&\colhead{Beam\tablenotemark{c}} \\
\colhead{}& \colhead{}& \colhead{}&
\colhead{(GHz)}&\colhead{}& \colhead{(min)}&\colhead{($\mu$Jy beam$^{-1}$)}
&\colhead{(mas)}}
\startdata
J0046+0104&1999FEB24&J0049+0237&4.99&LCP&55&107&$3.8\times 2.0$, PA $17^\circ$ \\
J0804+6459&1999FEB24&J0756+6347&1.67&LCP&154&71&$6.3\times 4.8$, PA $49^\circ$ \\
J0804+6459&1999FEB24&J0756+6347&4.99&LCP&235&67&$2.1\times 1.6$, PA $37^\circ$ \\
J1219+0638&2000JAN21&J1222+0413&1.42&LCP&125&77&$11.5\times 5.4$, PA $-2^\circ$ \\
J1219+0638&2000JAN21&J1222+0413&2.27&RCP&124&102&$7.0\times 3.3$, PA $-6^\circ$ \\
J1219+0638&2000JAN21&J1222+0413&4.99&LCP&152&76&$3.2\times 1.5$, PA $-8^\circ$ \\
J1316+0051\tablenotemark{d}&2000MAR12&J1256$-$0547&4.99&LCP&50&\nodata&\nodata \\
J1353+6345&2000FEB06&J1344+6606&1.42&LCP&135&90&$6.6\times 5.4$, PA $4^\circ$ \\
J1353+6345&2000FEB06&J1344+6606&2.27&RCP&129&95&$4.0\times 3.3$, PA $-9^\circ$ \\
J1353+6345&2000FEB06&J1344+6606&4.99&LCP&165&68&$1.8\times 1.5$, PA $-5^\circ$ \\
J1436+5847&2000MAR12&J1510+5702&1.67&LCP&165&79&$6.2\times 4.4$, PA $20^\circ$ \\
J1436+5847&2000MAR12&J1510+5702&4.99&LCP&228&66&$2.1\times 1.4$, PA $1^\circ$ \\
\enddata
\tablenotetext{a}{Polarization is denoted by ``LCP'' for Left Circular
Polarization and ``RCP'' for Right Circular Polarization.}
\tablenotetext{b}{The noise level quoted is that for the naturally
weighted data, providing the best possible sensitivity.}
\tablenotetext{c}{The beam sizes given are the full widths at half
maximum for data weighting optimizing the combination of resolution
and sensitivity.}
\tablenotetext{c}{Imaging of J1316+0051 failed, probably because of an
excessive distance of $8^\circ$ from the phase-referencing source.}
\label{tab:obs}
\end{deluxetable}

\clearpage

\begin{deluxetable}{lcccccc}
\tablecolumns{7}
\tablewidth{0pc}
\tablecaption{Measured Properties of RQQs}
\tablehead{
\colhead{(1)}& \colhead{(2)}& \colhead{(3)}& \colhead{(4)}&
\colhead{(5)}&\colhead{(6)}&\colhead{(7)} \\
\colhead{Name}&\colhead{Frequency}&\colhead{R.A. (J2000)\tablenotemark{a}}&
\colhead{Dec. (J2000)\tablenotemark{a}}& \colhead{$S_\nu$}&
\colhead{$P_\nu({\rm em})$\tablenotemark{b}}&\colhead{$T_{\rm B}$ (rest)\tablenotemark{c}} \\
\colhead{}&\colhead{(GHz)}&\colhead{(h m s)}&\colhead{(\arcdeg\ 
\arcmin\ \arcsec\ )}&\colhead{(mJy)}&\colhead{(W Hz$^{-1}$)}&
\colhead{(K)}}
\startdata
J0046+0104&4.99&00 46 13.5477&01 04 25.723&$4.7\pm 0.5$&$5.3\times 10^{25}$&$>5.6\times 10^8$ \\
J0804+6459&1.67&\nodata&\nodata&$24.9\pm 1.2$&$1.1\times 10^{24}$&\ \ \ $2.7\times 10^8$ \\
J0804+6459&4.99&08 04 30.4654&64 59 52.800&$6.5\pm 0.6$&$2.9\times 10^{23}$&\ \ \ $5.9\times 10^7$ \\
J1219+0638&1.42&\nodata&\nodata&$5.7\pm 0.3$&$1.6\times 10^{24}$&$>4.5\times 10^8$ \\
J1219+0638&2.27&\nodata&\nodata&$7.1\pm 0.4$&$2.0\times 10^{24}$&$>5.9\times 10^8$ \\
J1219+0638&4.99&12 19 20.9317&06 38 38.467&$6.4\pm 0.3$&$1.8\times 10^{24}$&$>5.1\times 10^8$ \\
J1353+6345&1.42&\nodata&\nodata&$21.3\pm 1.1$&$3.4\times 10^{23}$&$>2.4\times 10^9$ \\
J1353+6345&2.27&\nodata&\nodata&$14.5\pm 0.7$&$2.3\times 10^{23}$&$>1.7\times 10^9$ \\
J1353+6345&4.99&13 53 15.8307&63 45 45.686&$6.8\pm 0.4$&$1.1\times 10^{23}$&$>8.1\times 10^8$ \\
J1436+5847&1.67&\nodata&\nodata&$6.0\pm 0.3$&$1.2\times 10^{22}$&$>6.0\times 10^8$ \\
J1436+5847&4.99&14 36 22.0814&58 47 39.407&$3.1\pm 0.2$&$6.4\times 10^{21}$&$>3.1\times 10^8$ \\
\enddata
\tablenotetext{a}{Source positions are listed only at 4.99 GHz, since
these have the lowest errors, typically 1~mas or less in each coordinate.}
\tablenotetext{b}{Radio powers are quoted as the powers that were {\it emitted}
at the observed frequency.  K-corrections have been computed using the 
spectral indices between the lowest and highest VLBA frequencies, except for
J0046+0104, where the VLA spectral index listed in Table~\ref{tab:sample}
was used.}
\tablenotetext{c}{The listed brightness temperatures are in the rest frames
of the quasars, with the observed values multiplied by $(1+z)$. For
unresolved sources, the size upper limits are assumed to be half the beam
sizes.  For J0804+6459, the peak brightness temperatures are listed, 
using the beam size as the source size.}
\label{tab:tb}
\end{deluxetable}

\end{document}